# Analytic solutions of the quantum two-state problem in terms of the double-, bi- and tri-confluent Heun functions

**T.A. Shahverdyan**[1,2], **T.A. Ishkhanyan**[1,2,*], **A.E. Grigoryan**[1], **A.M. Ishkhanyan**[1]

[1]Institute for Physical Research, NAS of Armenia, Ashtarak, 0203 Armenia
[2]Moscow Institute of Physics and Technology, Dolgoprudny, 141700 Russia

We derive five classes of quantum time-dependent two-state models solvable in terms of the double-confluent Heun functions, five other classes solvable in terms of the bi-confluent Heun functions, and a class solvable in terms of the tri-confluent Heun functions. These classes generalize all the known families of two- or three-parametric models solvable in terms of the confluent hypergeometric functions to more general four-parametric classes involving three-parametric detuning modulation functions. The particular models derived describe different non-linear (parabolic, cubic, sinh, cosh, etc.) level-sweeping or level-glancing processes, double- or triple-level-crossing processes, as well as periodically repeated resonance-glancing or resonance-crossing processes. Finally, we show that more classes can be derived using the equations obeyed by certain functions involving the derivatives of the confluent Heun functions. We present an example of such a class for each of the three discussed confluent Heun equations.

## 1. Introduction

The few-state models suggest simple basic representations widely used for studying a number of phenomena in many branches of quantum physics [1-6]. This is a good starting approximation if during the interaction with an external field only a few of energy levels of a real quantum system are involved in transitions while the others are not. Within the framework of few-state representations, an important role has been played by the analytic solutions of the two-state problem, which have been extensively applied in collision physics, manipulation of atoms by laser fields, tunneling effects in complex biological systems, chemical dynamics, and, generally, in the theory of quantum non-adiabatic transitions [1-20].

In the present paper we discuss the solutions of the semi-classical time-dependent two-state problem in terms of the double-, bi- and tri-confluent Heun functions [21-23]. These are members of the Heun class of mathematical functions, which generalize many of the familiar special functions including the hypergeometric, Airy, Bessel, Mathieu functions, etc. We have recently studied the reduction of the two-state problem to the general and confluent Heun equations and have derived, respectively, thirty-five five-parametric [24] and fifteen four-parametric [25] classes of models allowing solutions in terms of these functions. Earlier, the bi-confluent Heun equation was used to extend the models solvable in terms of the confluent hypergeometric functions to five four-parametric classes of models solvable in terms of the bi-confluent Heun functions [26].

-----------------------------

[*]Corresponding author: E-mail: tishkhanyan@gmail.com

The approach we apply to find the field configurations for which the two-state problem is reduced to an equation having known analytic solution is based on the following general property of the solvable models. Consider the time-dependent Schrödinger equations defining the semiclassical time-dependent two-state problem. This is a system of coupled first-order differential equations for the probability amplitudes of the involved two quantum states $a_{1,2}(t)$ containing two arbitrary real functions of time, $U(t)$ and $\delta(t)$:

$$i\frac{da_1}{dt} = Ue^{-i\delta}a_2, \quad i\frac{da_2}{dt} = Ue^{+i\delta}a_1. \tag{1}$$

It is then checked that if the functions $a_{1,2}^*(z)$ solve this system rewritten for an auxiliary argument $z$ for some functions $U^*(z)$ and $\delta^*(z)$, then the functions $a_{1,2}(t) = a_{1,2}^*(z(t))$ solve the system (1) for the field-configuration defined as

$$U(t) = U^*(z)\frac{dz}{dt}, \quad \delta_t(t) = \delta_z^*(z)\frac{dz}{dt} \tag{2}$$

for arbitrary complex-valued function $z(t)$ [27-30].

This property allows one to group all the solvable models into separate classes, each of which includes the models that are derived from the same amplitude- and detuning-modulation functions $U^*(z)$ and $\delta_z^*(z)$ via transformations (2). Then, the search for the whole variety of models solvable in terms of a particular special function is reduced to the identification of the independent pairs $\{U^*, \delta_z^*\}$, referred to as the basic integrable models, for which the solution of the two-state problem is written in terms of this special function.

Here we consider the double-, bi-, and tri-confluent Heun functions, which are solutions of particular confluent modifications of the general Heun equation [31] arising by means of coalescence of some of its singular points [21-23]. The three equations under consideration can be written in the following form:

$$P(z)u_{zz} + (\gamma + \delta z + \varepsilon z^2)u_z + (\alpha z - q)u = 0, \tag{3}$$

where $P(z) = z^2$, $z$ and $1$ for the double-, bi- and tri-confluent Heun equations, respectively.

The technique to find two-state models for which the Schrödinger equations (1) are reduced to a target equation (i.e., in this case the double-, bi- and tri-confluent Heun equations (3)) is as follows. Consider the differential equation for $a_2$ derived from system (1) by elimination of $a_1$:

$$\frac{d^2 a_2}{dt^2} + \left(-i\delta_t - \frac{U_t}{U}\right)\frac{da_2}{dt} + U^2 a_2 = 0. \quad (4)$$

The transformation of the dependent variable $a_2 = \varphi(z)\,u(z)$ together with (2) reduces Eq. (4) to the following equation for the new dependent variable $u(z)$:

$$u_{zz} + \left(2\frac{\varphi_z}{\varphi} - i\delta_z^* - \frac{U_z^*}{U^*}\right) u_z + \left(\frac{\varphi_{zz}}{\varphi} + \left(-i\delta_z^* - \frac{U_z^*}{U^*}\right)\frac{\varphi_z}{\varphi} + U^{*2}\right) u = 0, \quad (5)$$

where and hereafter the lowercase Latin index denotes differentiation with respect to the corresponding variable. This equation becomes one of the discussed three types of confluent Heun equations (3) if

$$2\frac{\varphi_z}{\varphi} - i\delta_z^* - \frac{U_z^*}{U^*} = \frac{\gamma + \delta z + \varepsilon z^2}{P(z)} \quad (6)$$

and

$$\frac{\varphi_{zz}}{\varphi} + \left(-i\delta_z^* - \frac{U_z^*}{U^*}\right)\frac{\varphi_z}{\varphi} + U^{*2} = \frac{\alpha z - q}{P(z)} \quad (7)$$

with $P(z) = z^2$, $z$, 1. Though this is an underdetermined system of two nonlinear equations for three unknown functions, $U^*(z)$, $\delta^*(z)$ and $\varphi(z)$, the general solution of which is not known, however, many particular solutions can be found starting from an ansatz, which is guessed by inspecting the structure of the right-hand sides of Eqs. (6) and (7).

We have previously applied this approach to generalize the known models solvable in terms of the hypergeometric functions to six infinite three-parametric classes [28,29], as well as to generalize the models solvable in terms of the confluent hypergeometric functions to three infinite three-parametric classes [30]. As it was already mentioned above, we have recently discussed the solvability of the two-state problem in terms of the general and confluent Heun functions [24,25]. Thirty-five five-parametric and fifteen four-parametric classes of models allowing solutions in terms of these functions have been derived, a useful feature of which is the extension of the previously known detuning modulation functions - two-parametric at most - to functions involving more parameters. In the case of constant detuning this leads to two-peak symmetric or asymmetric pulses with controllable width [25], and, in the general case of variable detuning, it provides a variety of level-crossing models including symmetric and asymmetric models of non-linear sweeping through the resonance [32,33], level-glancing configurations [34-36], processes with two resonance-crossing time points [36] and multiple (periodically repeated) crossing models [37-40].

Here we follow the steps applied in [24-30] and derive five four-parametric basic models solvable in terms of the double-confluent Heun functions, five more such models solvable in terms of the bi-confluent Heun functions, and a model solvable in terms of the tri-confluent Heun functions. These models generalize all the known two- and three-parametric basic models solvable in terms of the confluent hypergeometric functions to more general four-parametric ones involving three-parametric detuning modulation functions. A subsequent transformation of the independent variable is further applied to generate different families of real field configurations with real Rabi frequency $U(t)$ and detuning $\delta(t)$. The derived models describe different non-linear-in-time (parabolic, cubic, sinh, cosh, etc.) level-sweeping and level-glancing, as well as double, triple and periodically repeated resonance-crossing processes.

## 2. Two-state models solvable in terms of the double-confluent Heun functions

The double-confluent Heun equation is written as [21-23]

$$\frac{d^2 u}{dz^2} + \left(\frac{\gamma}{z^2} + \frac{\delta}{z} + \varepsilon\right)\frac{du}{dz} + \frac{\alpha z - q}{z^2} u = 0. \tag{8}$$

Accordingly, Eqs. (6) and (7) are written as:

$$\frac{\gamma}{z^2} + \frac{\delta}{z} + \varepsilon = 2\frac{\varphi_z}{\varphi} - i\delta_z^* - \frac{U_z^*}{U^*}, \tag{9}$$

$$\frac{\alpha z - q}{z^2} = \frac{\varphi_{zz}}{\varphi} + \left(-i\delta_z^* - \frac{U_z^*}{U^*}\right)\frac{\varphi_z}{\varphi} + U^{*2}. \tag{10}$$

Examination of the structures of these equations suggests searching for their particular solutions in the following form:

$$\frac{\varphi_z}{\varphi} = \frac{\alpha_2}{z^2} + \frac{\alpha_1}{z} + \alpha_0 \Leftrightarrow \varphi = z^{\alpha_1} e^{\alpha_0 z - \frac{\alpha_2}{z}}, \tag{11}$$

$$\frac{U_z^*}{U^*} = \frac{k}{z} \Leftrightarrow U^* = U_0^* z^k, \tag{12}$$

$$\delta_z^* = \frac{\delta_2}{z^2} + \frac{\delta_1}{z} + \delta_0. \tag{13}$$

Multiplying now Eq. (10) by $z^4$ we get that, for arbitrary $\delta_{0,1,2}$, the product $U_0^{*2} z^{2k+4}$ should be a polynomial in $z$ of the fourth degree at most. Hence, $k$ is an integer or half-integer obeying the inequalities $0 \leq 2k+4 \leq 4$. This leads to five admissible cases of $k$, namely, $k = -2, -3/2, -1, -1/2, 0$, generating five respective classes of two-state models solvable in terms of the double-confluent Heun function. The amplitude modulation functions for these classes are given as

$$\frac{U^*}{U_0^*} = \frac{1}{z^2}, \ \frac{1}{z\sqrt{z}}, \ \frac{1}{z}, \ \frac{1}{\sqrt{z}}, \ 1. \qquad (14)$$

According to Eqs. (2), the actual field configurations, for which the solution of the two-state problem is written in terms of the double-confluent Heun function, are given as

$$U(t) = U_0^* z^k \frac{dz}{dt}, \qquad (15)$$

$$\delta_t(t) = \left(\frac{\delta_2}{z^2} + \frac{\delta_1}{z} + \delta_0\right)\frac{dz}{dt}, \qquad (16)$$

where $k = -2, -3/2, -1, -1/2, 0$ and the parameters $U_0^*$, $\delta_{0,1,2}$ are complex constants which should be chosen so that the functions $U(t)$ and $\delta(t)$ are real for the chosen complex-valued $z(t)$. Since these parameters are arbitrary, all the derived classes are 4-parametric in general.

The solution of the two-state problem (1) is explicitly written as

$$a_2 = z^{\alpha_1} e^{\alpha_0 z - \frac{\alpha_2}{z}} H_D(\gamma, \delta, \varepsilon; \alpha, q; z), \qquad (17)$$

where the parameters of the double-confluent Heun function $\gamma, \delta, \varepsilon, \alpha, q$ are given as

$$\gamma = 2\alpha_2 - i\delta_2, \ \delta = 2\alpha_1 - i\delta_1 - k, \ \varepsilon = 2\alpha_0 - i\delta_0, \qquad (18)$$

$$\alpha = \alpha_1 \varepsilon - \alpha_0 (k + i\delta_1) + Q'''(0)/6, \qquad (19)$$

$$q = \alpha_1(1 + k + i\delta_1 - \alpha_1) - \alpha_2 \varepsilon + i\alpha_0 \delta_2 - Q''(0)/2 \qquad (20)$$

with $Q(z) = U_0^{*2} z^{2k+4}$ and 

$$\alpha_0^2 - i\alpha_0 \delta_0 + Q^{(4)}(0)/4! = 0, \qquad (21)$$

$$\alpha_1 \gamma - \alpha_2(2 + k + i\delta_1) + Q'(0) = 0, \qquad (22)$$

$$\alpha_2^2 - i\alpha_2 \delta_2 + Q(0) = 0. \qquad (23)$$

Formal power-series expansions of the double-confluent Heun function $H_D$ are constructed using the substitution [21-23]:

$$H_D(\gamma, \delta, \varepsilon; \alpha, q; z) = e^{v_0 z - v_1/z} z^\mu \sum c_n z^n. \qquad (24)$$

However, the convergence radius of such a series is zero. Nevertheless, these expansions present asymptotic series that can be useful in both theoretical developments and practical applications [21-23]. For instance, if terminated, the series yield finite-sum closed-form exact solutions (quasi-polynomials).

The coefficients of the series (24) generally obey a five-term recurrence relation. However, the relation reduces to a three-term one if $v_1 = 0$ or $-\gamma$ and $v_0 = 0$ or $-\varepsilon$. For the simplest choice $v_0 = v_1 = 0$, the result is

$$R_n c_n + Q_{n-1} c_{n-1} + P_{n-2} c_{n-2} = 0 \tag{25}$$

with
$$R_n = \gamma(n+\mu), \tag{26}$$

$$Q_n = (n+\mu)(n+\mu+\delta-1) - q, \quad P_n = \alpha + \varepsilon(n+\mu). \tag{27}$$

If $\gamma \neq 0$, the series is left-hand side terminated if $\mu = 0$. It is then right-hand side terminated at some $n = N$ ($N = 1, 2, 3, ...$) if $\alpha + \varepsilon N = 0$ and $c_{N+1} = 0$. The last equation is a $(N+1)$th-order polynomial equation for the parameter $q$.

To present examples of field configurations generated by the basic models (13),(14) through Eqs. (15),(16), we consider the transformations $z(t) = e^t$ and $z(t) = e^{it}$.

With $z(t) = e^t$ and $U_0^* = U_0$, we have the families

$$U(t) = U_0 e^{(k+1)t}, \quad \delta_t(t) = \delta_0 e^t + \delta_1 + \delta_2 e^{-t}, \tag{28}$$

for which the detuning varies in such a way that the field frequency may cross the resonance $\delta_t = 0$ up to two times. For the class $k = -1$ the amplitude of the field is constant.

If $\delta_2 = \delta_0$, the field configuration for $k = -1$ is specified as

$$U(t) = U_0, \quad \delta_t(t) = \delta_1 + 2\delta_0 \cosh(t). \tag{29}$$

Here, non-crossing, level glancing and double crossing processes are possible depending on the parameters $\delta_{0,1}$ (Fig. 1). Level glancing takes place if $\delta_1 = -2\delta_0$.

If $\delta_2 = -\delta_0$, we have the following field configuration for $k = -1$:

$$U(t) = U_0, \quad \delta_t(t) = \delta_1 + 2\delta_0 \sinh(t), \tag{30}$$

which provides only one resonance crossing (Fig. 2). For nonzero $\delta_1$ the crossing is asymmetric in time, while if $\delta_1 = 0$, this is a symmetric quasi-linear-in-time level-crossing: $\delta_t \sim 2\delta_0 t$ at $t \to 0$ ( if the terminology of [32] is used, this is an example of super-linear crossing since at $t \to \infty$ the detuning diverges faster than the linear Landau-Zener detuning).

A solution of the two-state problem (1) for the configuration (28) corresponding to the choice $\alpha_{0,1,2} = 0$ ($\varphi(z) = 1$) is written as

$$a_2 = H_D(-i\delta_2, 1-i\delta_1, -i\delta_0; 0, -U_0^2; e^t). \tag{31}$$

The second independent solution is written using a different triad $\alpha_{0,1,2}$ (see Eqs. (21)-(23)).

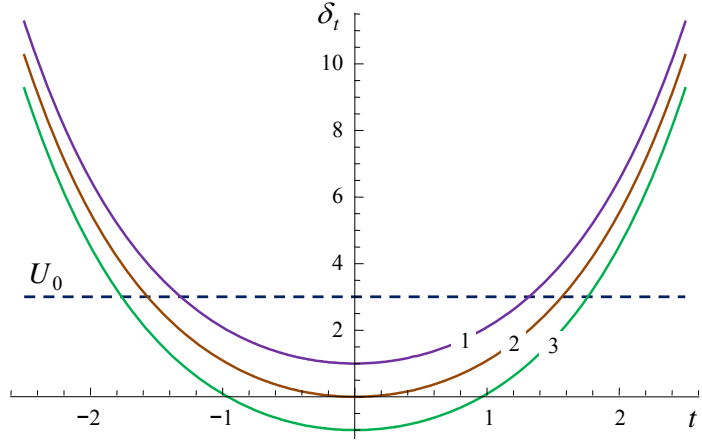

Fig. 1. Double-confluent Heun class $k = -1$, $z(t) = e^t$, $U_0 = 3$.
Detunings corresponding to $\delta_2 = \delta_0 = 1$ and $\delta_1 = -1; -2; -3$ (curves 1, 2, 3, respectively).

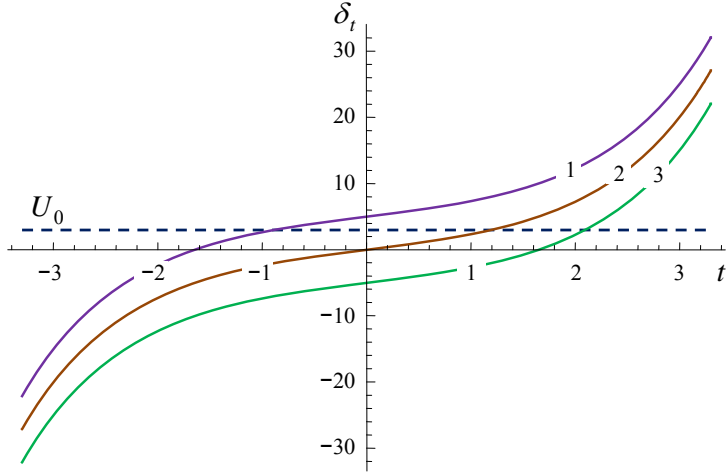

Fig. 2. Double-confluent Heun class $k = -1$, $z(t) = e^t$, $U_0 = 3$.
Detunings corresponding to $\delta_2 = -\delta_0 = -1$ and $\delta_1 = 5; 0; -5$ (curves 1, 2, 3, respectively).

A different level crossing model is obtained by the transformation $z(t) = e^{i(t-t_0)}$ ($t_0 = \text{const}$). Again, considering the class $k = -1$ and choosing $U_0^* = -iU_0$, $\delta_0 = \delta_2 = -i\Delta_2 / 2$ and $\delta_1 = -i\Delta_1$ we obtain the following field configuration

$$U(t) = U_0, \ \delta_t(t) = \Delta_1 + \Delta_2 \cos(t - t_0),  \qquad (32)$$

which provides periodically repeated level-glancing or resonance crossing processes (Fig. 3). For this case, a solution of the two-state problem corresponding to the choice $\alpha_{0,1,2} = 0$ is explicitly written as

$$a_2 = H_D(-\Delta_2 / 2, 1 - \Delta_1, -\Delta_2 / 2; 0, U_0^2; e^{i(t-t_0)}). \qquad (33)$$

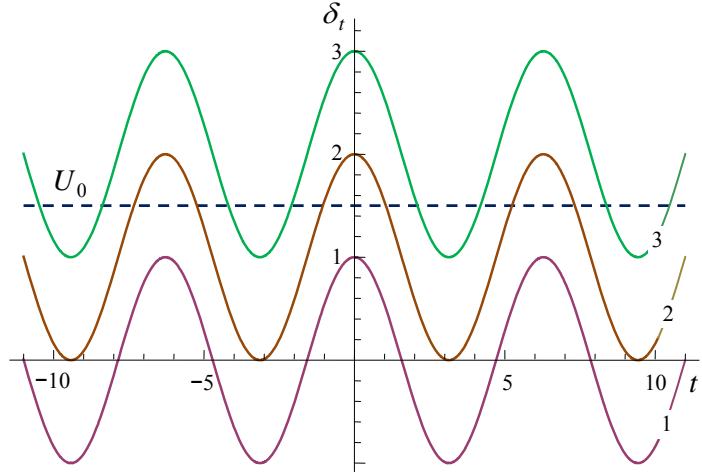

Fig. 3. Double-confluent Heun class $k = -1$, $z(t) = e^{i(t-t_0)}$, $U_0 = 1.5$.
Detunings corresponding to $t_0 = 0$, $\Delta_2 = 1$ and $\Delta_1 = 0; 1; 2$ (curves 1, 2, 3, respectively).

## 3. Two-state models solvable in terms of the bi-confluent Heun functions

The bi-confluent Heun equation is written as

$$\frac{d^2 u}{dz^2} + \left(\frac{\gamma}{z} + \delta + \varepsilon z\right)\frac{du}{dz} + \frac{\alpha z - q}{z} u = 0. \tag{34}$$

Hence, the equations (6) and (7) in this case are written as:

$$\frac{\gamma}{z} + \delta + \varepsilon z = 2\frac{\varphi_z}{\varphi} - i\delta_z^* - \frac{U_z^*}{U^*}, \tag{35}$$

$$\frac{\alpha z - q}{z} = \frac{\varphi_{zz}}{\varphi} + \left(-i\delta_z^* - \frac{U_z^*}{U^*}\right)\frac{\varphi_z}{\varphi} + U^{*2}. \tag{36}$$

We search for solutions of these equations in the following form:

$$\frac{\varphi_z}{\varphi} = \frac{\alpha_1}{z} + \alpha_0 + \alpha_2 z \Leftrightarrow \varphi = z^{\alpha_1} e^{\alpha_0 z + \frac{\alpha_2}{2} z^2}, \tag{37}$$

$$\frac{U_z^*}{U^*} = \frac{k}{z} \Leftrightarrow U^* = U_0^* z^k, \tag{38}$$

$$\delta_z^* = \frac{\delta_1}{z} + \delta_0 + \delta_2 z. \tag{39}$$

Multiplying Eq. (36) by $z^2$ we get that, for arbitrary $\delta_{0,1,2}$, the product $U_0^{*2} z^{2k+2}$ should be a polynomial in $z$ of the fourth degree at most. Hence, $k$ is an integer or half-integer obeying the inequalities $0 \le 2k + 2 \le 4$. This leads to five admissible cases of $k$, namely, $k = -1, -1/2, 0, 1/2, 1$, generating five classes of two-state models solvable in terms of the bi-confluent Heun functions. The amplitude modulation functions for these classes are

$$\frac{U^*}{U_0^*} = \frac{1}{z}, \quad \frac{1}{\sqrt{z}}, \quad 1, \quad \sqrt{z}, \quad z. \tag{40}$$

Thus, the field configurations, for which the solution of the two-state problem is written in terms of the bi-confluent Heun functions, are given as [26]

$$U(t) = U_0^* z^k \frac{dz}{dt}, \tag{41}$$

$$\delta_t(t) = \left(\frac{\delta_1}{z} + \delta_0 + \delta_2 z\right)\frac{dz}{dt} \tag{42}$$

with $k = -1, -1/2, 0, 1/2, 1$, and $U_0^*$, $\delta_{0,1,2}$ being complex constants which should be chosen so that the functions $U(t)$ and $\delta(t)$ are real for the chosen complex-valued $z(t)$. Since these parameters are arbitrary, the classes are four-parametric.

The solution of the initial two-state problem (1) is explicitly written as

$$a_2 = z^{\alpha_1} e^{\alpha_0 z + \frac{\alpha_2}{2} z^2} H_B(\gamma, \delta, \varepsilon; \alpha, q; z), \tag{43}$$

where the parameters of the bi-confluent Heun function $\gamma$, $\delta$, $\varepsilon$, $\alpha$, $q$ are given as

$$\gamma = 2\alpha_1 - i\delta_1 - k, \quad \delta = 2\alpha_0 - i\delta_0, \quad \varepsilon = 2\alpha_2 - i\delta_2 \tag{44}$$

$$\alpha = \alpha_0(\alpha_0 - i\delta_0) + \alpha_1(2\alpha_2 - i\delta_2) + \alpha_2(1 - k - i\delta_1) + Q''(0)/2, \tag{45}$$

$$q = \alpha_0(k + i\delta_1) - \alpha_1(2\alpha_0 - i\delta_0) - Q'(0) \tag{46}$$

with $Q(z) = U_0^{*2} z^{2k+2}$ and

$$\alpha_0 \varepsilon - i\alpha_2 \delta_0 + Q'''(0)/3! = 0, \tag{47}$$

$$\alpha_1^2 - \alpha_1(1 + k + i\delta_1) + Q(0) = 0, \tag{48}$$

$$\alpha_2^2 - i\alpha_2 \delta_2 + Q^{(4)}(0)/4! = 0. \tag{49}$$

The origin is a regular singular point of Eq. (34). Hence, the equation permits of a Frobenius power-series solution:

$$u = z^\mu \sum_{n=0}^{\infty} c_n z^n. \tag{50}$$

Substitution of Eq. (50) into Eq. (34) gives the following recurrence relation for the coefficients of the expansion:

$$R_n c_n + Q_{n-1} c_{n-1} + P_{n-2} c_{n-2} = 0, \tag{51}$$

where
$$R_n = (n+\mu)(n+\mu-1+\gamma), \qquad (52)$$

$$Q_n = \delta(n+\mu) - q, \quad P_n = \alpha + \varepsilon(n+\mu), \qquad (53)$$

where one should put $\mu = 0$ or $\mu = 1-\gamma$.

Unlike the above power series for the double-confluent Heun equation, this series is convergent everywhere in the complex $z$-plane. The series is right-hand side terminated at some $n = N = 1, 2, 3, ...$, thus producing finite-sum polynomial solutions, if $\alpha + \varepsilon(N+\mu) = 0$ and $c_{N+1} = 0$. The latter equation is a $(N+1)$th-order polynomial equation for $q$.

The classes (41),(42) include all the known confluent-hypergeometric level-crossing models. For instance, by setting $\delta_0 = 0$ and making the replacement $z \to \sqrt{z}$, it is verified that the classes with $k = -1, 0, 1$ become the classes by Landau–Zener-Majorana-Stückelberg [9-12], Nikitin [5,13] and Crothers [14], respectively.

In addition, the derived classes suggest several other interesting level-crossing models. One example is provided by the class $k = -1/2$. With specifications $U_0^* = U_0/2$, $\delta_0 = \Delta_1/2$, $\delta_1 = 0$, $\delta_2 = \Delta_2/2$ and transformation $z(t) = t^2$, we have a non-linear-in-time level-crossing model, namely, a constant-amplitude cubic-crossing model providing one or three crossings of the resonance (Fig. 4):

$$U(t) = U_0, \quad \delta_t(t) = \Delta_1 t + \Delta_2 t^3. \qquad (54)$$

A solution of the two-state problem for this model corresponding to the choice $\alpha_{1,2,3} = 0$ (see Eqs. (47)-(49)) is written as

$$a_2 = H_B\left(\frac{1}{2}, -\frac{i\Delta_1}{2}, -\frac{i\Delta_2}{2}; 0, -\frac{U_0^2}{4}; t^2\right). \qquad (55)$$

Another class, $k = +1/2$ with specifications $U_0^* = 3U_0/2$, $\delta_0 = \delta_1 = 0$, $\delta_2 = 3\Delta_2/2$ and transformation $z(t) = t^{2/3}$, presents an example of a constant-amplitude level-crossing model with infinitely fast sweeping through the resonance (Fig. 5):

$$U(t) = U_0, \quad \delta_t(t) = \Delta_2 t^{1/3}. \qquad (56)$$

A solution of the two-state problem for this model is written as (here we take $\alpha_{1,2} = 0$)

$$a_2 = e^{\delta t^{2/3}/2} H_B\left(-\frac{1}{2}, \delta, -\frac{3i\Delta_2}{2}; \frac{\delta^2}{4}, \frac{\delta}{4}; t^{2/3}\right), \qquad (57)$$

where $\delta = -3iU_0^2/\Delta_2$, and we assume $t > 0$.

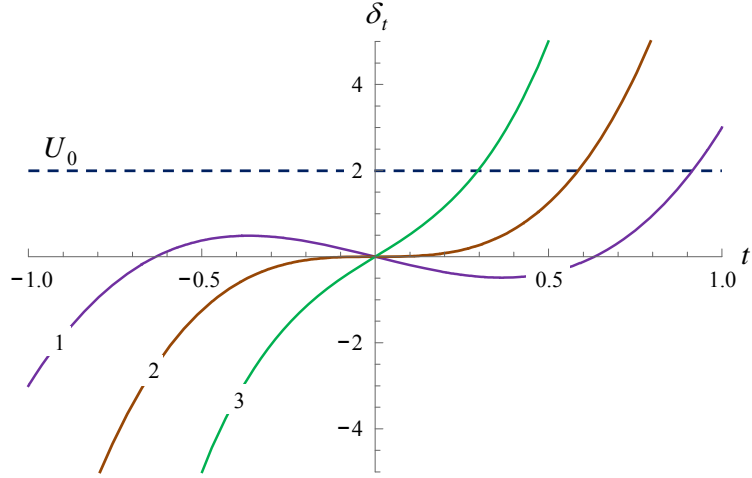

Fig. 4. Bi-confluent Heun class $k = -1/2$, $z(t) = t^2$, $U_0 = 2$.
Detunings corresponding to $(\Delta_1, \Delta_2) = (-2,5); (0,10); (5,20)$ (curves 1, 2, 3, respectively).

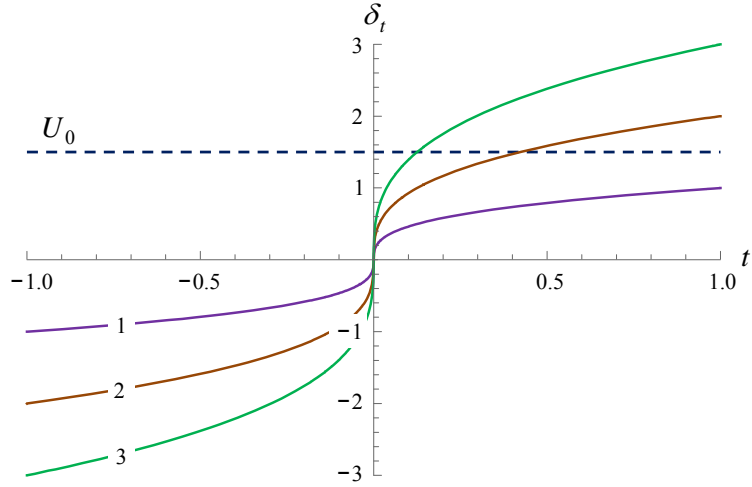

Fig. 5. Bi-confluent Heun class $k = +1/2$, $z(t) = t^{2/3}$, $U_0 = 1.5$.
Detunings corresponding to $\Delta_2 = 1; 2; 3$ (curves 1, 2, 3, respectively).

## 4. Two-state models solvable in terms of the tri-confluent Heun functions

The tri-confluent Heun equation is written as

$$\frac{d^2 u}{dz^2} + \left(\gamma + \delta z + \varepsilon z^2\right)\frac{du}{dz} + (\alpha z - q)u = 0. \tag{58}$$

Respectively, Eqs. (6) and (7) take the form:

$$\gamma + \delta z + \varepsilon z^2 = 2\frac{\varphi_z}{\varphi} - i\delta_z^* - \frac{U_z^*}{U^*}, \tag{59}$$

$$\alpha z - q = \frac{\varphi_{zz}}{\varphi} + \left(-i\delta_z^* - \frac{U_z^*}{U^*}\right)\frac{\varphi_z}{\varphi} + U^{*2}. \tag{60}$$

Inspecting the structure of Eq. (59), we search for solutions of these equations in the form:

$$\frac{\varphi_z}{\varphi} = \alpha_0 + \alpha_1 z + \alpha_2 z^2, \quad \delta_z^* = \delta_0 + \delta_1 z + \delta_2 z^2, \quad \frac{U_z^*}{U^*} = k_0 + k_1 z + k_2 z^2. \tag{61}$$

The second equation then immediately shows that there is only one choice: $k_{1,2,3} = 0$, so that $U^* = U_0^* = \text{const}$. This defines a four-parametric class with field configuration given as

$$U(t) = U_0^* \frac{dz}{dt}, \quad \delta_t(t) = (\delta_0 + \delta_1 z + \delta_2 z^2)\frac{dz}{dt}. \tag{62}$$

The solution of the initial two-state problem for this class is written as

$$a_2 = e^{\alpha_0 z + \frac{\alpha_1}{2}z^2 + \frac{\alpha_2}{3}z^3} H_T(\gamma, \delta, \varepsilon; \alpha, q; z), \tag{63}$$

where the involved parameters are given as

$$\alpha_{1,2,3} = \{0,0,0\}, \quad \{\gamma, \delta, \varepsilon, \alpha, q\} = \{-i\delta_0, -i\delta_1, -i\delta_2, 0, -U_0^{*2}\}, \tag{64}$$

or $\quad \alpha_{1,2,3} = \{i\delta_0, i\delta_1, i\delta_2\}, \quad \{\gamma, \delta, \varepsilon, \alpha, q\} = \{i\delta_0, i\delta_1, i\delta_2, 2i\delta_2, -U_0^{*2} - i\delta_1\}, \tag{65}$

The tri-confluent Heun equation (58) has only one singular point. This is an irregular singularity of rank 3, located at $z = \infty$. Since the origin is an ordinary point, the equation permits of a power-series solution of the form:

$$u = z^\mu \sum_{n=0}^{\infty} c_n z^n, \tag{66}$$

which is convergent everywhere. However, this time the recurrence relation for the coefficients of the expansion generally involves four terms:

$$S_n c_n + R_{n-1} c_{n-1} + Q_{n-2} c_{n-2} + P_{n-3} c_{n-3} = 0, \tag{67}$$

where $\quad S_n = (n+\mu)(n+\mu-1), \quad R_n = \gamma(n+\mu), \tag{68}$

$$Q_n = \delta(n+\mu) - q, \quad P_n = \alpha + \varepsilon(n+\mu). \tag{69}$$

Here, in order to have a consistent series, from two characteristic exponents, $\mu = 0, 1$, one should take the greater one, $\mu = 1$ (the other one leads to a logarithmic solution). The series is right-hand side terminated at some $n = N = 1, 2, 3, ...$ if $\alpha + \varepsilon(N + \mu) = 0$ and the parameters of the equation meet the conditions $c_{N+1} = c_{N+2} = 0$.

Concerning the point that the recurrence relation (67) involves four terms, the following remark is appropriate. As compared with the confluent hypergeometric classes, the field configuration (62) provides an extension only if $\delta_2 \neq 0$, which implies that $\varepsilon \neq 0$ (since $\varepsilon = \pm i\delta_2$, see Eqs. (64),(65)). However, in the case of non-zero $\varepsilon$ one always may achieve $\gamma = 0$ by shifting the origin: $z \to z - z_0$. Since then $R_n = 0$ vanishes for all $n$, we see that one can always reduce the recurrence relation (67) to one involving only three terms, however, non-successive.

Applying the transformation $z = \Delta(t - t_1)$ and using the specifications $U_0^* = U_0/\Delta$, $\delta_0 = 0$, $\delta_1 = (t_1 - t_2)/\Delta$, $\delta_2 = 1/\Delta^2$ we get a field configuration with constant amplitude and parabolic detuning (Fig. 6):

$$U(t) = U_0, \quad \delta_t(t) = \Delta(t - t_1)(t - t_2). \tag{70}$$

This is a non-crossing model (if $t_{1,2}$ have non-zero imaginary parts) or double-crossing model (if $t_{1,2}$ are real and not zero simultaneously), and it is a level-glancing model if $t_1 = t_2 = 0$.

A solution of the two-state problem for this model corresponding to $\alpha_{1,2,3} = 0$ is

$$a_2 = H_T\left(0, -\frac{i(t_1 - t_2)}{\Delta}, -\frac{i}{\Delta^2}; 0, -\frac{U_0^2}{\Delta^2}; \Delta(t - t_1)\right). \tag{71}$$

The second independent solution is given by Eq. (63) using the second set of the parameters given by Eq. (65).

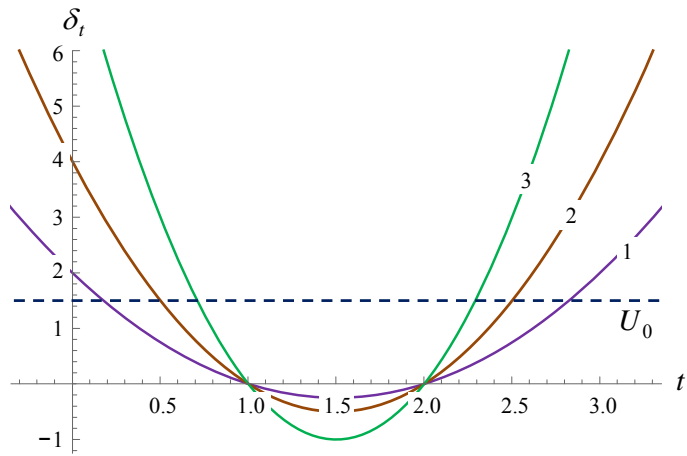

Fig. 6. Tri-confluent Heun class, $z = \Delta(t - t_1)$, $U_0 = 1.5$.
Detunings corresponding to $t_1 = 1$, $t_2 = 2$ and $\Delta = 1; 2; 4$ (curves 1, 2, 3, respectively).

## 5. Other classes solvable in terms of the Heun functions

The above classes do not cover all the field configurations, for which the solution of the two-state problem is written in terms of the Heun functions. Several other classes can be derived, e.g., if one considers the equations obeyed by the derivatives of the Heun functions. This is because the latter functions generally do not belong to the Heun class, but obey more complicated equations generally involving one more regular singular point [41-44].

Consider, for instance, the tri-confluent Heun equation (58). It is readily shown that the weighted first derivative of its solution $v(z) = e^{\gamma z} u_z$ obeys the equation

$$v_{zz} + \left(-\gamma + \delta z + \varepsilon z^2 - \frac{1}{z - z_0}\right) v_z + \frac{\Pi(z)}{z - z_0} v = 0, \qquad (72)$$

where $z_0 = q/\alpha$ and $\Pi(z)$ is the cubic polynomial

$$\Pi(z) = -\gamma \varepsilon z^3 + (\alpha + \varepsilon - \gamma \delta + z_0 \gamma \varepsilon) z^2 + z_0 (\gamma \delta - 2\alpha - 2\varepsilon) z + z_0 (q - \delta). \qquad (73)$$

It is seen that for non-zero $\alpha$ this equation has an additional regular singularity located at the point $z = z_0$. If $q = 0$ (note that in the case of non-zero $\alpha$ one always may achieve this by shifting the origin) and additionally $\alpha + \varepsilon = \gamma \delta$, the equation is simplified to

$$v_{zz} + \left(-\gamma + \delta z + \varepsilon z^2 - \frac{1}{z}\right) v_z - \gamma \varepsilon z^2 v = 0. \qquad (74)$$

Comparing now this equation with Eq. (4) rewritten for the variable $z$:

$$\frac{d^2 a_2^*}{dz^2} + \left(-i\delta_z^* - \frac{U_z^*}{U^*}\right) \frac{da_2^*}{dz} + U^{*2} a_2^* = 0, \qquad (75)$$

we see that $a_2^*(z) = v(z)$ if

$$\delta_z^* = \delta_0 + \delta_1 z + \delta_2 z^2, \; U^* = U_0^* z, \qquad (76)$$

and $\qquad \gamma = i\delta_0, \; \delta = -i\delta_1, \; \varepsilon = -i\delta_2, \; U_0^{*2} = -\gamma \varepsilon. \qquad (77)$

Hence, according to the property (2), the solution of the two-state problem (1) for the class of models given as (compare with Eq. (62))

$$U(t) = U_0^* z \frac{dz}{dt}, \; \delta_t(t) = (\delta_0 + \delta_1 z + \delta_2 z^2) \frac{dz}{dt} \qquad (78)$$

with arbitrary (complex-valued) function $z(t)$ is written in terms of the derivative of a tri-confluent Heun function (we recall that $q = 0$ and $\alpha + \varepsilon = \gamma \delta$):

$$a_2 = e^{i\delta_0 z} \frac{d}{dz} H_T(i\delta_0, -i\delta_1, -i\delta_2; i\delta_2 + \delta_0 \delta_1; 0; z). \qquad (79)$$

However, not all parameters here are independent. The last Eq. (77) imposes the constraint $U_0^{*2} = -\delta_0 \delta_2$. Thus, the class (78) is three-parametric.

Similar classes are readily derived using the derivatives of the bi-confluent and double confluent Heun functions. For instance, the function $v = z^\sigma u_z$, where $u(z)$ is a solution of the bi-confluent Heun equation (34), obeys the equation

$$v_{zz} + \left(\frac{\gamma + 1 - 2\sigma}{z} + \delta + \varepsilon z - \frac{1}{z - z_0}\right) v_z + \frac{(\alpha + \varepsilon - \varepsilon\sigma)(z - z_0)^2}{z^2} v = 0, \qquad (80)$$

where $z_0 = q/\alpha$, $\sigma = \gamma + \delta z_0 + \varepsilon z_0^2$ and $(\alpha + \varepsilon)z_0 = (\delta + 2\varepsilon z_0)\sigma$. Comparing this equation with Eq. (75) we immediately find the class

$$U(t) = U_0^* \frac{z - z_0}{z} \frac{dz}{dt}, \quad \delta_t(t) = \left(\frac{\delta_1}{z} + \delta_0 + \delta_2 z\right) \frac{dz}{dt}, \qquad (81)$$

for which the solution of the two-state problem is written as

$$a_2 = z^\sigma \frac{d}{dz} H_B(-i\delta_1 + 2\sigma, -i\delta_0, -i\delta_2; \alpha; \alpha z_0; z) \qquad (82)$$

with $\quad \sigma = i(\delta_1 + \delta_0 z_0 + \delta_2 z_0^2)$, $\alpha = i\delta_2(1 - 2\sigma) - i\delta_0 \sigma / z_0$, $U_0^{*2} = \alpha + \varepsilon - \varepsilon\sigma$. $\qquad$ (83)

Here, $\delta_{0,1,2}$ and $z_0$ are arbitrary real parameters, hence, this is a four-parametric class.

Constant-amplitude family of field configurations is achieved by the transformation

$$z(t) = -z_0 W(-e^{-t/z_0}/z_0), \qquad (84)$$

where $W$ is the Lambert product log function [45]. A corresponding family of chirped detunings is shown in Fig. 7.

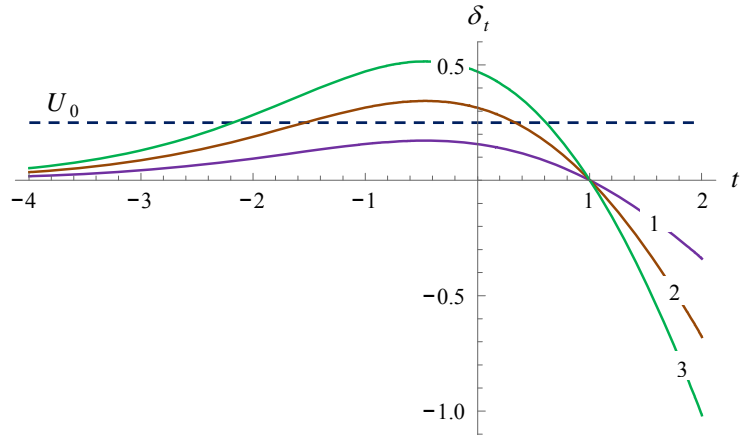

Fig. 7. Bi-confluent Heun class (81). Constant-amplitude family with $z_0 = -1$: $z = W(e^t)$, $\delta_1 = 0$, $\delta_2 = -\delta_0$. Detunings corresponding to $\delta_0 = 1; 2; 3$ (curves 1, 2, 3, respectively).

Similarly, the function $v = e^{-\sigma/z} u_z$ with $u(z)$ being a solution of the double-confluent Heun equation (8) with $\alpha = -\varepsilon$ and $q = \delta/2$ obeys the equation

$$v_{zz} + \left(\frac{\gamma - 2\sigma}{z^2} + \frac{\delta + 2}{z} + \varepsilon - \frac{1}{z - z_0}\right) v_z - \frac{\varepsilon \sigma (z - z_0)^2}{z^4} v = 0, \tag{85}$$

where $z_0 = q/\alpha$ and $\sigma = \gamma - \delta^2/(4\varepsilon)$. Comparing this equation with Eq. (75) we immediately find the three-parametric class

$$U(t) = U_0^* \frac{z - z_0}{z^2} \frac{dz}{dt}, \quad \delta_t(t) = \left(\frac{\delta_2}{z^2} + \frac{\delta_1}{z} + \delta_0\right) \frac{dz}{dt} \tag{86}$$

with $U_0^{*2} = -\varepsilon \sigma$, for which the solution of the two-state problem is written as

$$a_2 = e^{-\sigma/z} \frac{d}{dz} H_D(-i\delta_2 + 2\sigma, -i\delta_1, -i\delta_0; i\delta_0; -i\delta_1/2; z). \tag{87}$$

Because of the constraint $U_0^{*2} = -\varepsilon \sigma$, this is a three-parametric class. The constant-amplitude family is now achieved by the transformation (compare with Eq. (84))

$$z(t) = -\frac{z_0}{W(-z_0 e^{-t})} \tag{88}$$

with $z_0 = -\delta_1/(2\delta_0)$. A one-parametric family of constant-amplitude field configurations describing asymmetrically chirped detunings is shown in Fig. 8. Of course, other families, e.g., describing constant-detuning pulses, can be constructed by an appropriate choice of the transformation $z(t)$ and the involved parameters.

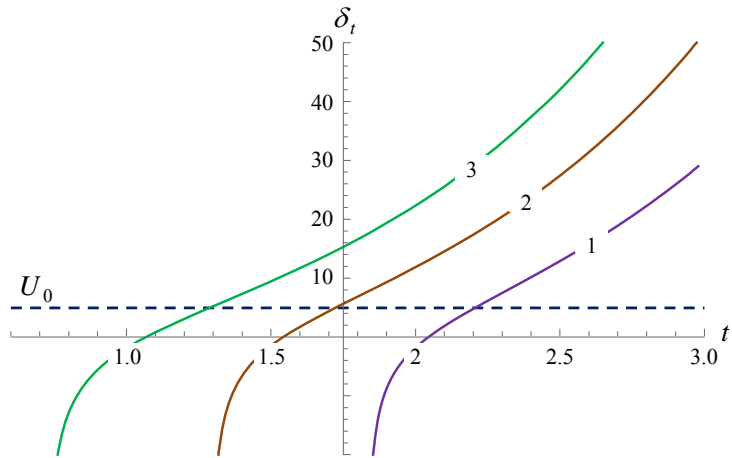

Fig. 8. Double-confluent Heun class (86). Constant-amplitude family with $z(t) = -z_0/W(-z_0 e^{-t})$, $z_0 = -\delta_1/(2\delta_0)$, $\delta_1 = -2\sqrt{25 - \delta_0^2}$, $\delta_2 = -\delta_0$. Detunings corresponding to $\delta_0 = 2; 3; 4$ (curves 1, 2, 3, respectively).

## 6. Discussion

Thus, we have presented five classes of quantum time-dependent two-state models solvable in terms of the double-confluent Heun functions, five other classes solvable in terms of the bi-confluent Heun functions, and a class solvable in terms of the tri-confluent Heun functions. All the derived classes are four-parametric. In the case of constant-amplitude field configuration, the models describe different non-linear (parabolic, cubic, sinh, cosh, etc.) level-sweeping or level-glancing, double- or triple-level-crossing processes. Models describing periodically repeated resonance-glancing or resonance-crossing are also possible.

We have indicated that many other classes can be derived using equations obeyed by certain functions involving the derivatives of the Heun functions. We have presented an example of such a class for each of the three confluent Heun equations. It should be noted that one cannot construct similar additional classes if the hypergeometric equations alone are considered. This is because the derivatives of the hypergeometric functions are again hypergeometric functions, while the derivatives of the Heun functions generally do not belong to the Heun class, but obey more complicated equations generally involving one more regular singular point [41-44].

The Heun functions used are solutions of corresponding confluent modifications of the general Heun equation, derived due to coalescence of some of its singular points. For some values of the involved parameters, each of these three equations is reduced to the confluent hypergeometric equation either directly or by a transformation of the dependent or independent variable. The above classes then reproduce the three known three-parametric classes solvable in terms of the confluent hypergeometric functions [30]

For instance, the double-confluent Heun equation is directly reduced to the confluent hypergeometric equation if $\gamma = q = 0$. We then note that in this case the system (18)-(23) becomes over-determined and permits of a solution only for the classes $k = -1, -1/2, 0$. Similarly, the bi-confluent Heun equation is directly reduced to the confluent hypergeometric equation if $\varepsilon = \alpha = 0$. In this case the system (44)-(49) becomes over-determined and it is readily checked that the solution exists only for the classes with $k = -1, -1/2, 0$. In both cases we have $\alpha_2 = \delta_2 = 0$, so that these classes exactly reproduce the three three-parametric confluent hypergeometric classes [30].

However, the tri-confluent Heun equation is not directly reduced to the confluent hypergeometric equation. If $\varepsilon = 0$, one needs a transformation of both independent and dependent variables, while if $\varepsilon \neq 0$ and $\gamma = \delta = 0$ the cubic transformation of the independent variable $z \to -\varepsilon(z-z_0)^3/3$ will suffice. It is then checked that the tri-confluent class (62) in each of these cases reproduces a confluent hypergeometric class (e.g., obviously, in the case $\gamma=\varepsilon=\alpha=0$ we immediately have the Landau-Zener model). Another example of reproducing the confluent hypergeometric classes was mentioned above for the bi-confluent Heun equation, achieved by the change $z \to \sqrt{z}$. The latter case is of particular interest since in this case the reproduction is achieved within the classes with $k=-1,0,1$. Hence, in addition to the classes $k=-1,-1/2,0$, the bi-confluent class $k=+1$ also includes a confluent hypergeometric subclass. Finally, it can be checked that this is the case for all other classes. Then, the conclusion is that all the presented double-, bi- and tri-confluent classes of two-state models, each in its own manner, present different generalizations of the prototype confluent hypergeometric families of models.

We would like to conclude by a brief discussion of the solutions of the confluent Heun equations. These are complicated functions, the theory of which still needs development. We have presented above the basic power-series solutions of the considered confluent Heun equations. However, in practical applications, especially, if non-linear extensions are discussed [46], one may need more advanced techniques. In this case, one may try expansions in terms of functions other than mere powers, e.g., in terms of familiar special functions such as the Kummer and Tricomi confluent hypergeometric functions [21-23,47-50], Bessel functions [50,51], Gauss hypergeometric functions [21-23,51], Coulomb wave functions [52-54], incomplete Gamma- and Beta-functions [41,55], etc. Expansions in terms of higher transcendental functions [56], e.g., in terms of the Goursat and the Appell generalized hypergeometric functions, are also possible [57,58]. In several cases these expansions may provide exact finite-sum solutions.

**Acknowledgments**

This research has been conducted within the scope of the International Associated Laboratory (CNRS-France and SCS-Armenia) IRMAS. The research has received funding from the European Union Seventh Framework Programme (FP7/2007-2013) under grant agreement No. 295025 – IPERA. The work has been supported by the Armenian State Committee of Science (SCS Grant No. 13RB-052). T.A. Ishkhanyan acknowledges the Dynasty Foundation for a Stipend for Physics Students as well as the support from SPIE through a 2015 Optics and Photonics Education Scholarship.